\documentclass[a4,12pt]{article}
\usepackage[english]{babel}
\usepackage{amssymb,amsmath}
\usepackage{authblk}
\input{epsf}

\newcommand{\be}{\begin{equation}}
\newcommand{\ee}{\end{equation}}

\newcommand{\symbolbold}{\bf}
\renewcommand{\vec}{\bf}

\begin{document}

\title{A solution to the problems of cusps and rotation
curves in dark matter halos in the cosmological standard model}

\author{A. G. Doroshkevich\thanks{Electronic address: \texttt{dorr@asc.rssi.ru}}, V. N. Lukash, E. V. Mikheeva}
\affil{\it Astro Space Center of Lebedev Physical Institute, Russian
Academy of Sciences, Profsoyuznaya st. 84/32, Moscow, 117997,
Russia}

\date{}

\maketitle

\begin{abstract}
We discuss various aspects of the inner structure formation in
virialized dark matter (DM) halos that form as primordial density
inhomogeneities evolve in the cosmological standard model. The main
focus is on the study of central cusps/cores and of the profiles of
DM halo rotation curves, problems that reveal disagreements among
the theory, numerical simulations, and observations. A method that
was developed by the authors to describe equilibrium DM systems is
presented, which allows investigating these complex nonlinear
structures analytically and relating density distribution profiles
within a halo both to the parameters of the initial small-scale
inhomogeneity field and to the nonlinear relaxation characteristics
of gravitationally compressed matter. It is shown that cosmological
random motions of matter `heat up' the DM particles in collapsing
halos, suppressing cusp-like density profiles within developing
halos, facilitating the formation of DM cores in galaxies, and
providing an explanation for the difference between observed and
simulated galactic rotation curves. The analytic conclusions
obtained within this approach can be confirmed by the N-body model
simulation once improved spatial resolution is achieved for central
halo regions.
\end{abstract}

\newpage

\tableofcontents

\section{Indroduction}

This review is the second of two (see~\cite{1_Lukash2011}) devoted
to the formation of the large-scale structure of the Universe and
problems of virialized dark matter (DM) halos. We do not attempt to
discuss all related problems of the physical cosmology and focus in
fact on DM evolution in the cosmological standard model (CSM). The
discussion is based on the original papers of the authors and
follows the chapters of book~\cite{2_Lukash2010}. Here, we present a
more detailed discussion of these problems using new observational
data and a comparison with the theory without invoking detailed
analytic calculations (theoretical results are described in the
appendices; see~\cite{2_Lukash2010} for the proofs).

\section{The essence and history of the problem}

Explaining the internal structure formation of virialized DM halos,
which are studied both observationally and using analytic and
numerical simulations, is one of the key cosmological issues.

Despite cosmologists' significant efforts, many aspects of halo
formation remain unclear. In particular, it is unclear whether the
universal density profile~\cite{3_Navarro1995}-\cite{5_Navarro1997}
(hereafter, the Navarro-Frenk-White (NFW) profile), discovered in a
wide range of masses and sizes of DM halo formation models, is truly
universal. The NFW profile has been found to correspond well to the
observed density profiles in galaxy
clusters~\cite{6_Pointecouteau2005}, but at the same time it
deviates from the observed galactic profiles in many respects
(see~\cite{7_Tasitsiomi2004,8_Hayashi2004}). This means that many
physical factors controlling galaxy formation are not fully taken
into account in numerical models.

Numerical calculations of the N-body problem, which have been widely
used for many years, are presently a powerful tool to study the
formation of the nonlinear structure of the Universe. But numerical
models have many restrictions, which stimulates the development of
analytic and approximate methods that can reveal and qualitatively
characterize the effect of the main physical processes responsible
for halo formation. One of the key processes that are difficult to
model numerically is the relaxation of matter contracted in a
protohalo. To analyze this process, the cosmological small-scale
matter velocity and density fields should be taken into account.

In the framework of the widely used spherically symmetric model of
the collapse, direct relations between the properties of the initial
density perturbations and virialized halos are given by the
Press-Schechter relations or their extensions (see~
\cite{9_Peebles1967}-\cite{20_Manrique2003}). In this approach, the
collapse is described as a consecutive fall of spherical shells,
with their subsequent relaxation and redistribution inside the halo.
Each spherical layer is characterized by the mass and turn-around
radius (the moment of `decoupling' from the Hubble expansion), which
can be related to the spectrum of initial density perturbations, for
example, using the smoothing of the density perturbation field by
filters of different scales (see~\cite{17_Bond1996}). This approach
can be generalized to include the angular momentum of the shells,
external tidal forces, an external density distribution,
etc.~\cite{21_White1992}-\cite{26_Williams2004}. Another description
of halo formation,which is not based on the quasispherical collapse
but uses the Zeldovich approximation, is proposed
in~\cite{27_Demianski2004}.

It is well known that the formation of any dense object means the
relaxation of contracted matter. In the spherical collapse model,
this is the so-called violent relaxation process, which provides the
energy and mass redistribution of the compressed matter, the removal
of the excess energy from the system, and the halo contraction. The
simplest example of the violent relaxation is given by the collapse
of a homogeneous nonrotating DM cloud of a finite size
(see~\cite{28_Fillmore1984}-\cite{30_Gurevich1995}). The angular
momentum does not stop the violent
relaxation~\cite{21_White1992}-\cite{23_Nusser2001}, but does weaken
it (see, for example,~\cite{26_Williams2004,31_Avila-Reese1998}).

In contrast to such methodically transparent results, numerical
calculations show that to describe the halo formation process, it is
insufficient to consider the fall of only diffusive matter
(spherically symmetric or anisotropic). In numerical models, the
formation of a halo is accompanied by the formation of random
systems of low-mass high-density subhalos and their subsequent
coalescence in the main protohalo. This process of hierarchical
merger provides the mass transfer inside the main halo, in which
small subhalos are `stripped' by the tidal interaction and dynamical
friction with the surrounding matter
(see~\cite{32_Tormen1997}-\cite{35_Dekel2003}). Both processes --
the fall of diffusive matter and the merger of satellite subhalos --
play similar roles in the halo formation.

The results of numerical calculations suggest that the merger of
early formed high-density subhalos can stimulate the formation and
later increase in the central density in the main halo, and the
violent relaxation here has a smaller effect. For example, even one
merger of subhalos with comparable masses can strongly change the
internal structure of the whole halo~\cite{36_Gao2004}. The merged
subhalos supply the common halo with their kinetic energies and
angular momenta, and hence the destruction and stripping of subhalos
inside the main halo become the principal constituents of the
relaxation process. The interaction of subhalos with the main halo,
other subhalos, and diffusive matter redistributes the energy in the
main halo and changes its internal structure. The efficiency of
relaxation in numerical calculations is comparable with estimates
obtained in the simplest spherically symmetric collapse models
(see~\cite{31_Avila-Reese1998,37_Voglis1995,38_Zaroubi1996}).

The relaxation is needed to provide the energy removal from the main
halo, but at the same time it leads to the chaotization of the
collected material and largely determines the density profile
unification in the relaxed halos. Nevertheless, in both the
hierarchical and spherical halo formation models, the merger of
subhalos (or spherical shells) is not purely chaotic and is
controlled by the initial density and velocity distributions. During
the relaxation, the initial conditions do not smear out completely
and are partially conserved. As we show below, both halo mass
distribution and their internal structure depend on the initial
perturbation spectrum.

It is well known that initial conditions can dramatically affect the
evolution of dynamical systems. The difference between the cold and
hot dark matter cosmological scenarios provides an example. In the
context of halo formation, the possible influence of thermal
velocities and small-scale initial perturbations were discussed
in~\cite{26_Williams2004,27_Demianski2004,39_Avila-Reese2001,40_Demianski1999}.
Here, we significantly improve the description of this effect by
using statistical characteristics of spatial density and velocity
distributions inside a collapsed DM cloud~\cite{41_Mikheeva2007}.
This approach allows estimating the background entropy of the
relaxed matter, which is related to the initial small-scale velocity
perturbations, and its dependence on the mass of halos formed. The
effect of factors such as the angular momentum of collapsing matter
and nonlinear destruction of low-mass subhalos is not considered
here. Due to the random character, these factors can only increase
the background entropy and its deviation from the mean value. Our
calculations yield the minimum level of the background entropy,
which, nevertheless, can change the internal structure of the
emerging halos and significantly suppress the central DM cusp
formation inside them.

The use of the entropy approach in describing relaxed halos allows
the determination of the total entropy (initial entropy and that
acquired during nonlinear relaxation) of DM particles in the halo.
In contrast to the kinetic approach, our method does not appeal to
the halo particle distribution and does not describe the relaxation
history. But it correctly reproduces any distribution of matter
inside a relaxed spherical halo and distinguishes between the
properties of this distribution that are due to the initial
conditions and those that are related to the relaxation process
itself. Such a calculation is possible due to the property of
entropy as the most stable characteristic of matter, changing only
during irreversible processes.

For both spherical and hierarchical halo formation models, it is
possible to estimate the entropy growth during matter
relaxation~\cite{42_Taylor2001}. For the spherical collapse, this
estimate can be done analytically (see~\cite{28_Fillmore1984}). The
hierarchical relaxation process is more complicated and is not
described analytically, and we therefore use the results of
numerical calculations to estimate the entropy growth there.

In the observed gravitationally bound objects, the spatial
distributions of DM, baryons, and luminous matter are different. In
the DM-dominated Universe, it is reasonable to regard the DM halo
formation problem as the first approximation to a more complicated
problem of the formation of gravitationally bound systems. Here, we
consider only the DM dynamics in the CSM framework, ignoring the
effect of dark baryons and luminous matter. Nevertheless, in some
cases where the effect of the baryon component on the internal
structure of a galaxy can be significant (see,
e.g.,~\cite{43_Ostriker2001}), this issue requires further studies.

\section{Dark matter halo}

We assume DM to consist of nonrelativistic massive particles that
interact with each other and other particles only gravitationally.
This matter is called `dark' because it is invisible -- it has a
nonbaryonic nature and does not interact with light.\footnote{Matter
can be weakly visible if its particles weakly interact with light,
for example, if they annihilate to emit photons in the regions with
high DM density in galactic centers.} However, DM can be studied
using dynamical methods, because it is clumped and produces spatial
gradients of the gravitational potential that influence the motion
of visible bodies (galaxies, stars, gas), the state of baryons (hot
gas), and light ray deflection (gravitational lensing).

The mean DM density in the Universe is five times higher than the
density of cosmological baryons, which is why nonrelativistic dark
particles control the process of gravitational clustering and of the
evolution of the inhomogeneous part of the gravitational potential
of the Universe. The density contrast of a weakly inhomogeneous
initial spatial DM distribution increases with time. Because this
matter is cold, the pressure gradients there are small and cannot
prevent the development of gravitational instability.

In regions with enhanced density, the Friedman expansion rate slows
down, and at some instant the DM expansion stops and turns into
collapse. During dynamical contraction and subsequent oscillations
of matter flows, processes of violent collisionless gravitational
relaxation occur: particles move in a variable gravitational
potential localized in space, such that some particles ($\sim 10\%$)
are expelled from the system and carry away the excess positive
energy. The remaining energy of the clump is redistributed inside
the system, and a gravitationally bound DM object called a halo is
formed.

The observed halos (galaxies, their groups, and clusters) are
relaxed systems of particles that are gravitationally bound in all
three spatial directions. The upper limits of the halo sizes and
masses are a few Mpc and $10^{15} M_\odot$. We recall that the mass
of matter in the Universe in a sphere with a radius $R = 10$ Mpc is
$M_{10}\simeq 1.6\cdot 10^{14} M_\odot$ [see Section 8, formula
(21)]. Here and below in numerical estimates, we assume the CSM
parameters
\begin{align}
\Omega_{\rm m}&\equiv\frac{\rho_{\rm m}}{\rho_{\rm c}}=1-\Omega_{\rm
E} \simeq 0.3 \,,\quad \rho_{\rm c}=\frac{3H_0^{\,2}}{8\pi G}\simeq
10^{-29} \mbox{ g }\mbox{sm}^{-3}\,, \nonumber \\
H_0&\simeq
70\mbox{ km }\mbox{s}^{-1}\mbox{ Mpc}^{-1}\,, \label{omega}
\end{align}
where $\rho_{\rm m}$ is the cosmological DM density, $\rho_{\rm c}$
is the critical density in the Universe, $\Omega_{\rm E}$ is the
cosmological density of dark energy expressed in units of the
critical density, $H_0$ is the Hubble constant, and $G$ is the
gravitational constant.

We consider the simplest equilibrium conditions of self-gravitating
DM systems. The distribution function of non-relativistic particles
in a spherically symmetric halo depends on the radial distance to
the halo center $r$ and moduli of the radial and transverse particle
momenta. In spherical coordinates, the stress-energy tensor of
matter takes the diagonal form
\be
T_\mu^\nu = \rho\; {\rm diag} \left(1, -\sigma_r^2, -\sigma_t^2,
-\sigma_t^2 \right),
\ee
where $\rho=\rho(r)$ is the halo density profile, and $\sigma_r(r)$
and $\sigma_t(r)$ are the radial and transverse velocity dispersions
of particles. In the Newtonian limit, the identity
$T_{\mu\,;\nu}^\nu=0$ implies the hydrostatic equilibrium equation
for collisionless particles:
\be\label{erp}
-\frac{1}{\rho r^2} \frac{d}{dr} \left(\rho\, r^2\sigma_r^2\right)
+\frac{2}{r} \, \sigma_t^2 = \frac{d\Phi}{d r} =
\frac{GM(r)}{r^2}\,,
\ee
where the gravitational potential $\Phi$, defined up to an additive
constant, and the mass of the system $M$, which depends on $r$, are
related to density as
\be
\Delta\Phi \equiv \frac{1}{r^2} \frac{d}{dr}\left(r^2 \frac{d\Phi
}{dr}\right) = 4\pi G\rho \,,\;\quad M(r) = 4\pi\int_0^r \rho r^2
dr\,.
\ee

Inside the halo, DM particles move along different orbits, from
radial to circular ones. Each orbit is characterized by the energy
and angular momentum vector, which are integrals of motion in a
static spherically symmetric field. The density distribution at the
halo center can be conveniently characterized by a power-law
function,
\be\label{alp}
\rho\propto r^{-\alpha}
\ee
with the exponent (slope) $\alpha={\rm const}\in (0;\, 2.5)$. The
particle distribution function depends on the evolutionary history
of the halo. Numerical experiments and observations show that for
most of the spherical halos, the particle velocity distribution is
nearly isotropic, although deviations from isotropy can sometimes be
as high as 20--30\% (see, for example,~\cite{44_Kravtsov2009}).

Large-mass halos do not have enough time to form in the Universe.
For example, systems collapsed in one or two dimensions, which are
called walls and filaments, are typical nonlinear elements of the
large-scale structure of the Universe. On average in the Universe,
the density contrast on scales exceeding dozen Megaparsecs remains
smaller than unity ($\vert\delta\rho/\rho\vert < 1$), and we can
consider only regions with an increased or decreased (in comparison
with the mean cosmological) density of matter.

\section{Entropy of particles in the halo}

We consider an isotropic distribution of DM particles with
$\sigma_r\!=\sigma_t\!=\sigma$ in more detail. The halo equilibrium
is determined by the balance of the effective pressure gradient of
nonrelativistic matter,
\be
p=nT=\rho\sigma^2
\ee
and the gravitation produced by the total mass $M = M(r)$ [see (3)],
\be\label{isot}
\frac{dp}{\rho\,dr} = - \frac{GM(r)}{r^2}\,,
\ee
where radial functions $n=\rho/m$ and $T$ are the density and
effective temperature of particles with the mass $m = {\rm const}$.
By virtue of the equivalence principle, $m$ cannot be determined
from gravitational equations because particles are moving along
geodesics irrespective of their masses. The measurable variables are
the density $\rho$ and the particle velocity dispersion $\sigma$.

From the hydrostatic equilibrium equation, it is possible to find
the radial matter density profile in the halo using the known
velocity dispersion distribution $\sigma (r)$; vice versa, using the
density distribution, we can reconstruct the velocity dispersion
law. For the adiabatic density distribution, $\sigma \propto
n^{1/3}$, where the coefficient of proportionality depends on the
entropy distribution. In analogy with an ideal gas, we introduce the
entropy function $E = E(r)$ of the virialized halo
as~\cite{2_Lukash2010, 41_Mikheeva2007}
\be\label{FE}
{\rm E} = \sigma^2 \left(\frac{m_{\rm p}}{\rho}\right)^{2/3} \propto
\frac{T}{n^{2/3}} = \frac{p}{n^{5/3}}\,,
\ee
where $m_{\rm p}$ is the mass of a proton to which the halo particle
mass $m$ is normalized.

The function $E(r)$ is the measure of the total entropy of halo
particles acquired during the full halo formation
history.\footnote{We stress that $E$ is a function of entropy and
not the entropy itself. Here, we do not consider general issues of
the applicability of the notion of entropy (coarse-grained entropy,
see footnote~3) to collisionless particles, by defining the function
$E(r)$ only for stationary equilibrium systems with isotropic
particle velocity distributions. We recall that in definition (8),
$\sigma^2=\bar{v^2}$ , where $v$ is the one-dimensional peculiar
velocity of dark matter particles at a point $r$ (the bar means
averaging over the velocity space).} It mainly includes the
background entropy determined by the initial small-scale flows and
matter inhomogeneities in the protohalo, and the acquired entropy
generated during collisionless and hierarchical relaxation of matter
at the nonlinear stage of the halo formation.

\section{Isothermal sphere}

Modeling the internal structure of observed halos frequently
involves the isothermal sphere approximation, where the velocity
dispersion can be assumed constant and independent of the radius:
\be
\sigma (r)=\sigma_0={\rm const} \,,\qquad
\kappa_0\equiv\frac{4\pi G}{\sigma_0^2}=\frac{4\pi Gm}{T}={\rm
const} \,.
\ee
This approximation is in good agreement with observational data in a
restricted range of scales (Fig.~1).

\begin{figure}[t]
\epsfxsize=100 mm \centerline{\epsfbox{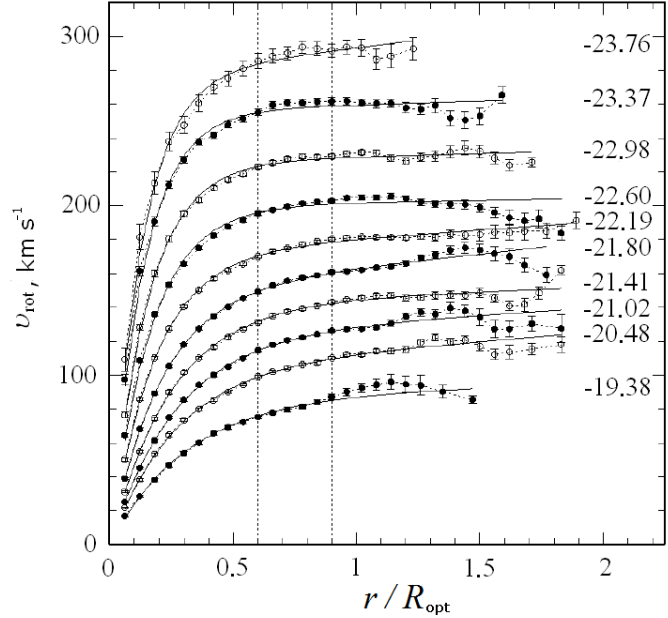}}
\caption{Rotation curves for 2155 galaxies
(from~\cite{45_Catinella2006}). The curves, which are divided into
several groups in accordance with the galaxy luminosities, are
characterized by the absolute stellar magnitude in the I-filter
(shown to the right). $R_{\rm opt}$ is the `optical' radius of
galaxies, inside which 83\% of the integral emission flux is
contained.} \label{rot1}
\end{figure}

After multiplying Eqn (7) by $r^2$ and differentiating with respect
to $r$, we obtain the equation for the density profile in an
isothermal sphere:
\be\label{zsot}
\frac{\;\;\;d}{\rho\, r^2 dr}\left(\frac{r^2 d\rho}{\rho\; dr}
\right)= - \kappa_0\,.
\ee
This nonlinear equation contains an attractor -- a particular
solution that is an attracting separatrix for dynamical trajectories
(10). The general solution has the form
\be\label{iso}
\rho(r) = \left\{
\begin{array}{ll}
\displaystyle \frac{\rho_0}{1+r^2/(2r_0^2)}\;,\qquad &r<r_0 \equiv
\displaystyle
 \sqrt{\frac{3}{\kappa_0\rho_0}} \\
 \\
\displaystyle \frac{2}{\kappa_0 r^2}\;, &r>r_0\quad
(\mbox{attractor})\,,
\end{array}
\right.
\ee
where the parameter $\rho_0$ determines the central density of the
halo core and $r_0$ is the characteristic size of the core.
Irrespective of the value of $\rho_0$, the density profile and mass
of the halo at $r > r_0$ are determined only by the velocity
dispersion of particles:
\be
M = \frac{8\pi r}{\kappa_0} \simeq 7\cdot 10^{12} \left(
\frac{\sigma}{ 300\;\mbox{km}\;\mbox{c}^{-1}} \right)^2
\left(\frac{r}{200\; \mbox{kpc}} \right) M_\odot\,.
\ee
For the characteristic values of $\sigma$ and $r$, we obtain the
typical mass of nonlinear dark matter halos in the observed Universe
-- that of galaxy groups. We recall that the total mass of the Local
Group, which includes the Milky Way and the Andromeda Nebula, is
$2\cdot 10^{12}\; M_\odot$.

We have derived Eqns (3) and (7) for DM by assuming its dominance;
however, these equations can be used to describe hydrostatic
equilibrium of any matter (DM, stars, gas) in the gravitational
field of the total mass $M(r)$. For this, the corresponding $p$ and
$\rho$ should be substituted in the left-hand side of Eqns (3) and
(7). For example, we can rewrite the halo mass in terms of the
effective gas temperature $T_{\rm gas}$ in the equilibrium:
\be
M = \frac{2 \,T_{\rm gas} \,r}{G\mu_{\rm gas}} \simeq 7 \cdot
10^{12} \left( \frac{T_{\rm gas}}{1 \, \mbox{keV}} \right) \, \left(
\frac{r}{200 \, \mbox{kpc}} \right) M_\odot\,,
\ee
where $\mu_{\rm gas} \simeq m_{\rm p} = 1\, \mbox{GeV}$ is the
molecular weight of the gas. Such a hot gas, which has been present
in galaxy clusters for several billion years, is observed by X-ray
telescopes, which allows reconstructing the gravitational potential
distribution and estimating the total mass of clusters (see,
e.g.,~\cite{46_Vikhlinin2009}).

We return to solution (11). The core density $\rho\simeq\rho_0$ does
not relate to the gravity of matter: it depends only on the initial
cosmological conditions that determined the halo formation history.
The size of the core is determined by the product of the particle
velocity and the dynamical time of the central density:
$$
r_0 \simeq\frac{\sigma_0}{2\sqrt{G\rho_0}}\,.
$$
Here, $r_0$ is the radius at which the matter self-gravity becomes
significant. At $r > r_0$, the dark matter gravity restructures the
inner halo such that the rotation velocities of particles cease to
depend on the radius:
\be\label{frq}
v_{\rm rot}(> r_0) = \sqrt{\frac{GM}{r}}=\sqrt 2 \,\sigma_0 = {\rm
const}\,.
\ee

\begin{figure}[t]
\epsfxsize=100 mm \centerline{\epsfbox{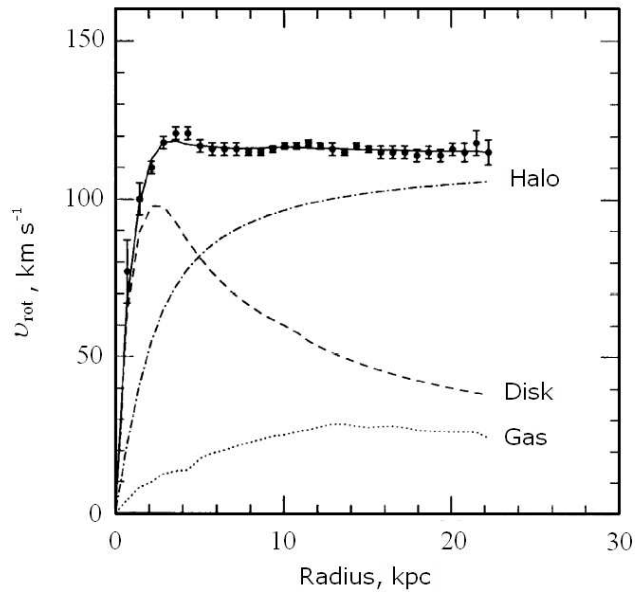}}
\caption{The rotation curve of galaxy NGC 6503 (data
from~\cite{47_Freese2008}). The dark matter halo contribution is
shown by the dashed-dotted line.} \label{rotcurves}
\end{figure}

Such velocity distributions are realized in many gravitationally
bound cosmological objects. For example, flat rotation curves are
observed in spiral galaxies (Fig.~2): circular velocities of stars
and gas first increase as the radius increases and then become
constant or change very slowly. The DM mass increasingly contributes
at large distances and maintains the flat part of the rotation curve
over about ten dynamical scales.

Dwarf galaxies with low surface brightness (Fig.~3) provide another
example. There, DM dominates starting from small radii, and a linear
increase in the circular rotation velocity is observed, suggesting a
small variation of the DM density in the central core [see (18)]. We
see that the initial increase of rotation curves also corresponds to
isothermal sphere (11):
\be
v_{\rm rot}(< r_0) = \sqrt{\frac{GM}{r}}=\frac{\sigma_0 r}{r_0}\,.
\ee
The fact that the inner circular velocities are smaller than the
particle velocities in the halo ($v_{\rm rot}<\sigma_0$) suggests
that the halo particles freely, by inertia, move over the region $r
\sim r_0$ with a constant velocity and do not feel gravity.

\begin{figure}[t]
\epsfxsize=70 mm \centerline{\epsfbox{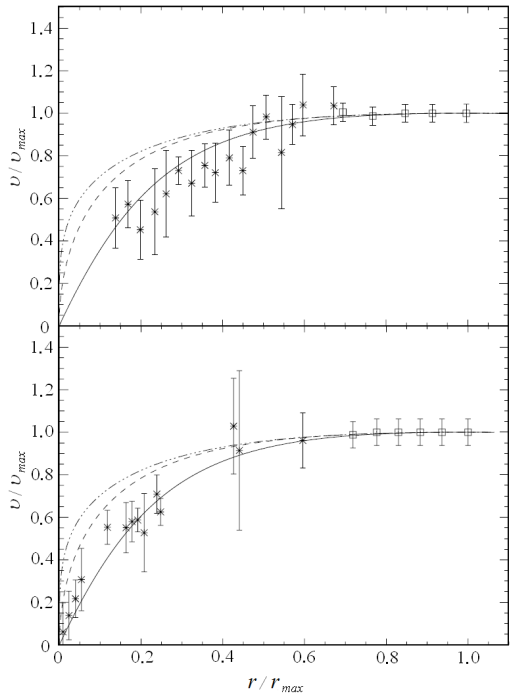}}
\caption{The circular rotation velocity as a function of radius in
low-brightness surface galaxies (a) LSB F571-8 and (b) LSB F583-1.
The solid curve shows the Burkert approximation, the dashed curve
shows the NFW approximation, and the dashed-dotted line shows the
Moore profile (data from~\cite{48_Marchesini2002}).} \label{ris3p10}
\end{figure}

To within the observational accuracy, isothermal sphere distribution
(11) can be approximated by the simple formula
\be\label{isots}
\rho\simeq\frac{\rho_0}{1+x^2}\,,\qquad x=\frac{r}{r_0}\,.
\ee
The flat part of the circular velocity, corresponding to the linear
mass increase $M\propto r$ [see (14)], is observed within a finite
dynamic range. The density distribution in such a system can be
conveniently described by the profile
\be\label{cisots}
\rho\propto\frac{1}{\left(1+x^2\right)\left(r+r_s\right)}\,,
\ee
where $r_s = {\rm const}$ is a characteristic cut-off parameter. In
the central part at $r < r_s$, the profile is close to that of the
isothermal sphere in (16); but at large distances, the density
decreases $\propto r^{-3}$ and circular velocities slowly decrease.
The circular velocity reaches a maximum in the region with the local
density slope exponent is $\alpha\simeq 2$.

\section{Internal halo structure}

The radial dependence of the circular velocity is related to the
density distribution inside the halo. Figure~3 demonstrates rotation
curves of dwarf galaxies that can be approximated well by
phenomenological profile (17) with $r_s = r_0$, which is referred to
as the Burkert approximation~\cite{49_Burkert1995}:
\be\label{burk}
\rho\propto\frac{1}{\left(1+x^2\right)\left(r+r_0\right)}\,.
\ee
Similar examples of galactic cores with slowly changing density are
typical for small galaxies. So far they have not been reproduced in
numerical calculations of structure formation.

The nonlinear relaxation of a contracting cloud leads to a diverging
central density in the forming halo. Analytic models of the
collapse~\cite{28_Fillmore1984,29_Gurevich1988} consider the
gravitational contraction of a spherical cloud at rest with a smooth
density distribution. After the relaxation, a singular halo density
profile is formed with the slope exponent $\alpha\sim 1.6 - 2$. This
result, which is also confirmed by numerical
calculations~\cite{50_Moore1994}-\cite{52_Diemand2005}, is closely
related to the choice of the initial state of the contracting mass,
namely, to the absence of random small-scale velocities and an
almost spherically symmetric density distribution. This example very
well illustrates the transformation of the kinetic energy of the
contracting cloud into thermal energy of the halo during violent
relaxation.

More realistic models of the collapse with subsequent DM relaxation
appear in numerical modeling of the evolution of a system of
gravitationally interacting N bodies ($N\sim 10^{10}$). Here, random
initial particle velocities, the anisotropy of the contraction, and
the successive merger of the main halo and its satellites are taken
into account. The mean density profile of halos formed in these
calculations is described by the NFW
approximation~\cite{3_Navarro1995}-\cite{5_Navarro1997}, which
depends on a single parameter $r_s$ and has the form
\be\label{NFW}
\rho\propto\frac{1}{r\left(r+r_s\right)^2}\,.
\ee
At $r < r_s$, the dependence of the mean density profile on $r$
tends asymptotically to a cusp-like power law with the slope $\alpha
= 1$.

\begin{figure}[t]
\epsfxsize=100 mm \centerline{\epsfbox{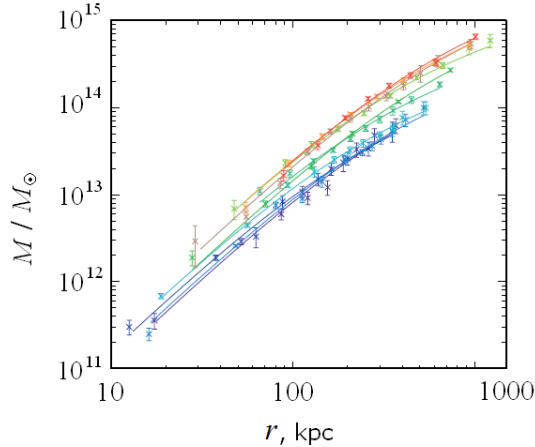}}
\caption{Dark matter mass distribution in galaxy clusters according
to the data in~\cite{6_Pointecouteau2005}.} \label{ris3p11}
\end{figure}

\begin{figure}[t]
\epsfxsize=100 mm \centerline{\epsfbox{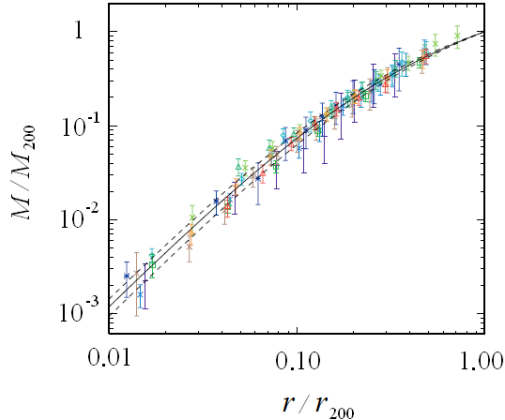}}
\caption{The curves shown in Fig.~4 in variables normalized to the
mass and radius of the sphere inside which the mean density is 200
times as high as the matter density in the Universe. The solid line
shows the NFW approximation.} \label{pointe2}
\end{figure}

The NFW profile obtained in numerical $N$-body models well describes
the density distribution in galaxy clusters (Fig.~4, 5). It should
be remembered, however, that this profile is obtained by averaging
density distributions of many halos, and deviations of individual
profiles from the mean one can exceed 20\% (Fig.~6). We also note
that technical limitations restrict the dynamical range of model
scales, and the properties of halos with low and moderate masses are
not properly reproduced in the calculations. Numerical constraints
also restrict the size of the central region resolved in numerical
models.

\begin{figure}[t]
\epsfxsize=100 mm \centerline{\epsfbox{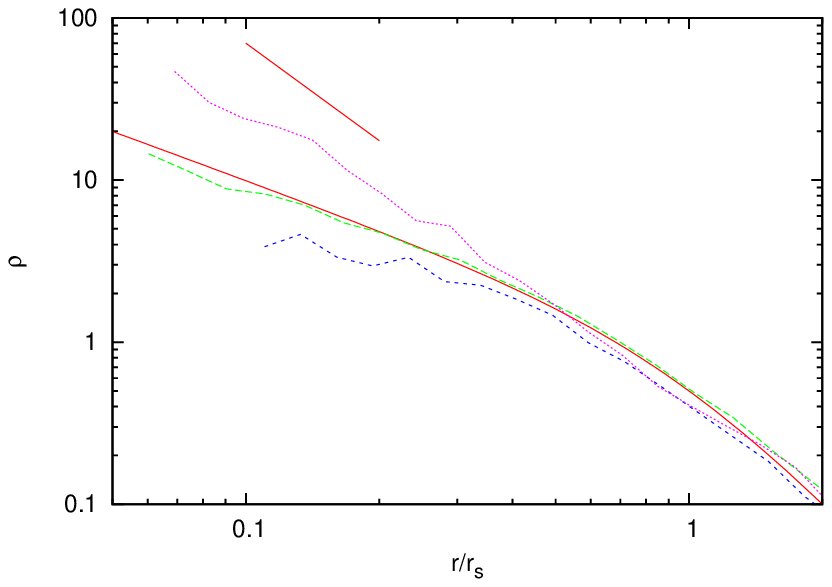}}
\caption{The density profiles of three halos with masses $10^{12} -
10^{13}$ obtained in numerical models~\cite{53_Pilipenko2009}. The
solid curve in the center shows the NFW profile. The short line in
the upper part of the figure corresponds to the slope $\alpha=2$.}
\label{newpilipenko}
\end{figure}

In numerical models, the DM halo central density tends to infinity,
but the mass remains finite and tidal forces are insignificant. Such
an internal structure was called the halo central cusp. Thus,
singular cusps are formed in numerical models, but these structures
are not observed in the Universe.

Figure~7 demonstrates the rapid density increase in a DM halo that
appears in numerical models. The comparison with observed galactic
profiles is shown in Fig.~8. The question of whether DM cusps exist
in reality is important not only for understanding galaxy formation;
it also can be connected with the physics of DM particles. For
example, one of the methods to search for possible DM particle
annihilation assumes that central cusps do exist in galactic
centers.

\begin{figure}[t]
\epsfxsize=100 mm \centerline{\epsfbox{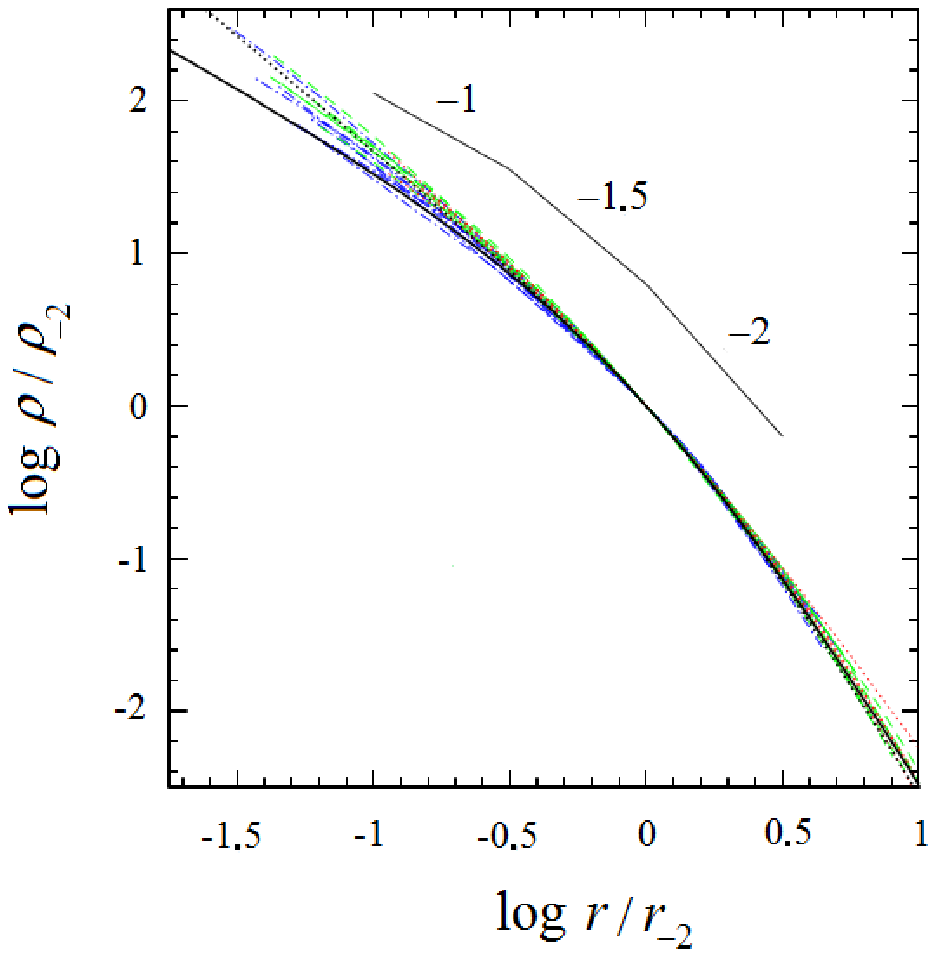}}
\caption{(See color version at www.ufn.ru). The mean halo density
profiles in numerical models~\cite{54_Navarro2003}. Dwarf galaxies
are shown in red, intermediate mass galaxies are shown in green, and
galaxy clusters are shown in blue. Numbers show the power-law slope
exponents $d\ln\rho/d\ln r$; the normalization corresponds to the
slope point $-2$.} \label{newris3p5}
\end{figure}

\begin{figure}[t]
\epsfxsize=100 mm \centerline{\epsfbox{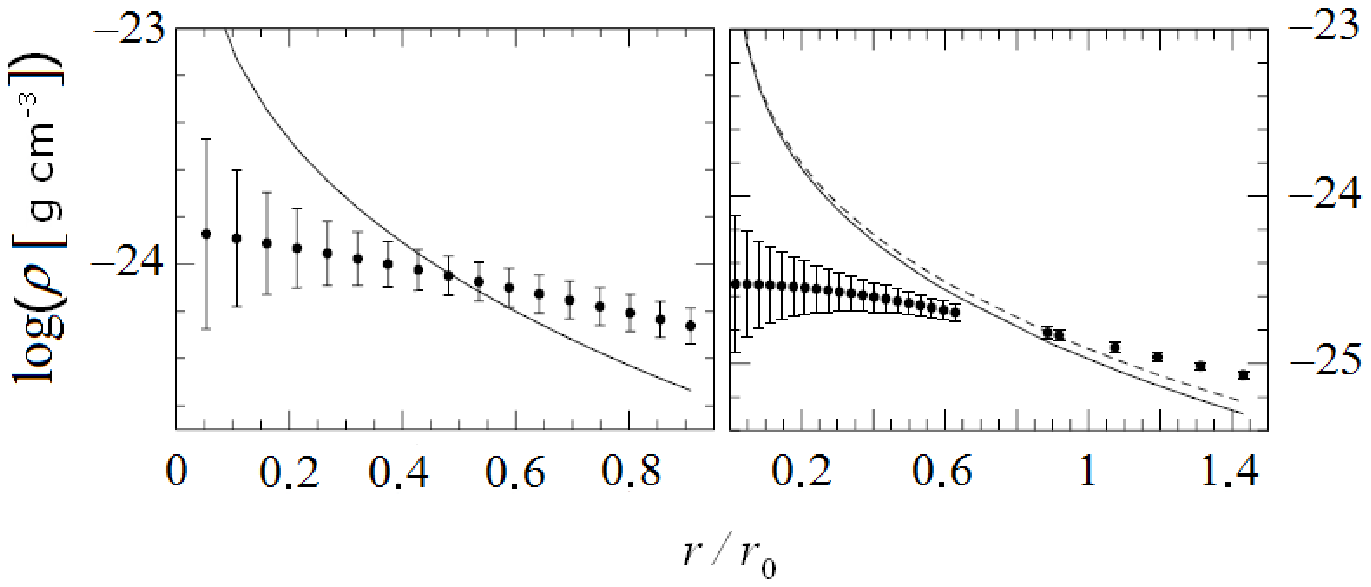}}
\caption{Demonstration of the cusp problem in galactic DM halos
(data from~\cite{55_Gentile2007}). The symbols show the observed
galactic density profiles in (a) DDO 47 and (b) ESO 287-G13, the
curves show the mean density profiles obtained in numerical
modeling.} \label{newris3p6}
\end{figure}

To summarize, the comparison of the observed galaxies with numerical
models leads to the following conclusions.

(1) Numerical models reproduce the formation of massive objects, and
the density profile of galaxy clusters is in good agreement with
calculations.

(2) Burkert profile (18) is mainly observed in low- and
moderate-mass galaxies dominated by DM. The formation of such
galaxies is related to the initial perturbations of comparatively
moderate scales. In the central regions of more massive spiral
galaxies, the DM and baryon densities are comparable and it is
difficult to separate them.

(3) These issues are closely related to the number count of our
Galaxy satellites. Until recently, the number of the observed
satellites was about two times less than predicted by numerical
models~\cite{44_Kravtsov2009}. However, many very faint satellites
have been discovered recently, and their number even exceeds the
prediction of numerical models
(see~\cite{56_Karachentsev2009}-\cite{60_Maccio2010}). This shows
that the properties of low-mass objects need to be studied further.

\section{What is needed to form cusps}

Do singular cusps that appear in numerical experiments actually form
in Nature? To answer this question, it is necessary to understand
the structure of cusps and what their formation conditions should
be.

As we discerned from the isothermal sphere example, DM particles
freely intersect the core region $r < r_0$ and do not stay there.
This conclusion is valid for much more numerous halos with a small
central mass ($\alpha < 1$), because the gravitational force tends
to zero at small radii:
\be\label{gf}
\frac{GM\left(r\right)}{r^2}\propto r^{1-\alpha}\rightarrow 0\,
\qquad\mbox{as}\qquad r\rightarrow 0\,.
\ee
Conversely, more massive cusps ($\alpha > 1$) gravitationally
influence DM particles and keep them inside the halo. Clearly,
velocities of the kept particles must be sufficiently low for the
particles to remain in the cusp region. Therefore, to form a cusp,
it is necessary to collect many cold particles in its center. At
least two questions arise.
\begin{itemize}
\item Where can cold particles in the perturbation field be found?
\item How can cold particles be transported to the central part of
the halo without being heated?
\end{itemize}

Because the fraction of energy removed during halo formation does
not exceed 10-15\% and the degree of matter contraction during
relaxation is finite, both these questions should be first of all
addressed by numerical programs for N-body calculations: where is
the `cooler' of DM particles and why is their `warming up' by
small-scale cosmological motions of matter not taken into account in
calculations?

We know that in the absence of small-scale perturbations, cusps are
formed in models of quasispherical collapse (see Section 6). In
these analytic calculations, all particles are initially focused
toward the center (by the chosen initial conditions), and a
significant fraction of the particles, by crossing the center, cools
due to violent relaxation and remains inside the cusp. But in real
cosmological conditions, DM particles are `defocused' by small-scale
perturbations and hence do not pass through the center. In addition,
the fall of DM clumps and their merging `stir up' the central region
and hamper effective particle cooling.

Of course, the quantitative characteristics of these mutually
opposite processes must be calculated and checked numerically;
however, modern numerical modeling is not perfect and needs to be
improved.

First of all, it is required to control DM motions in a wide range
of scales, from large cosmological scales on which the initial
particle distributions are given, to very small ones, which allows
the inner regions of the forming halos to be resolved. In modern
numerical models, the mass resolution, as a rule, does not exceed
$10^{7}\,M_\odot$, which is insufficient to describe the structure
of DM cores. Due to the restricted resolution in numerical models,
the initial particle distributions have no small-scale
perturbations, i.e., there is an effective cut-off of the
perturbation spectrum at short wavelengths. The consecutive
resolution of selected volumes by the addition of `new points'
inside them at the late stage of the evolution (the so-called
adaptive method) does not help either, because this procedure does
not add necessary information on the initial small-scale field of
cosmological perturbations.

A more sophisticated method is sometimes used that involves
consecutive recalculation of the model upon adding new particles in
the future dense halo regions. This procedure enables small-scale
perturbations to be included in the calculation in restricted
volumes. But such a recalculation complicates the problem and
strongly increases the computation time. The problem of cusps has
not yet been solved. Perhaps this example shows a limitation of the
CSM, in which DM particles are considered to be cold. In the CSM
framework, we cannot resolve the contradiction between observational
data and the results of numerical calculations. The point is that
cusps emerge during relaxation of the initially cold matter, in
which large-scale particle flows self-intersect to form caustics,
from which a cusp is later formed. But the real DM can be initially
warm with small random particle velocities. In that case, caustics
smear out, and cores with a smoother density distribution are formed
in halo centers instead of cusps.

Warm DM can be regarded as an additional parameter of the
cosmological model. The mass of warm DM particles that is required
to solve the problem of cusps must be not large ($\sim 10$ keV),
which allows the particles to preserve the residual thermal
velocities associated with the hot phase of evolution in the early
universe. Another possibility of the initial `heating up' of DM
particles can be related to the presence of excess power in the
perturbation spectrum at small wavelengths. We recall in this
connection that the spectrum is fairly well determined on scales
above several Mpc from observations of the cosmic microwave
background and large-scale galaxy distribution. At the same time, at
small scales, the spectrum of initial perturbations can deviate from
a power law, which can in turn affect the structure of small
galaxies.

These assumptions, however, go beyond the CSM framework. Our
approach is different. We believe that there are currently no strong
grounds to complicate the fundamental model, and the problem can
have a simpler solution, which does not require modification of the
CSM. The influence of small-scale velocity and density perturbations
on the internal halo structure can be studied analytically using
entropy function (8). The use of entropy instead of density is
motivated by the fact that for most particles, the entropy increases
during the halo formation and relaxation and is an integral
characteristic of the entire halo formation history. Evaporation of
particles during relaxation decreases the entropy, but this effect
is rather small.

By definition, DM particles are initially cold, i.e., their entropy
is zero. On the other hand, there are random small-scale flows and
matter clumps, which are tidally destructed and scattered when
approaching each other inside collapsing protohalos to form folds
and intersections, thus heating DM particles and chaoticizing their
velocities before the violent relaxation begins. Therefore, during
the first contraction, the small-scale velocity perturbations in the
protohalo are transformed into chaotic motions of DM particles. This
process can be described in terms of coarse-grained entropy, in
analogy with the description of the ideal
gas~\cite{41_Mikheeva2007}.

\section{Ensemble of protohalos}

To introduce the entropy characteristic of particles in a protohalo,
the notion of an ensemble is needed. It can be introduced as a
subsample of the field of linear perturbations embracing all
spatially bound regions that collapse by the time corresponding to
the current redshift $z$. A schematic comparison with an ideal gas
is shown in Fig.~9.

\begin{figure}[t]
\epsfxsize=100 mm \centerline{\epsfbox{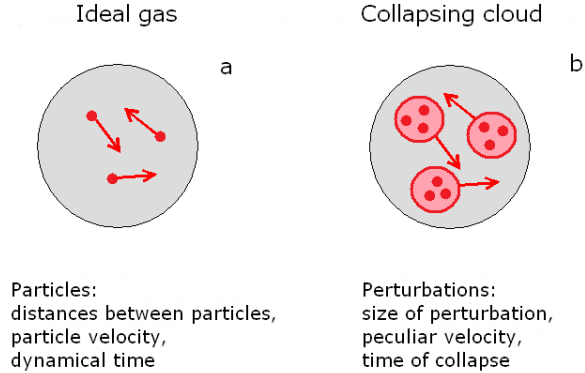}}
\caption{Motion of particles in (a) an ideal gas and (b) a DM
protohalo~\cite{2_Lukash2010}. The entropy of particles in the
region of size $r$ is determined on the dynamical time scale
coincident with the collapse time: $r/v\simeq H^{-1}$.}
\label{ris342}
\end{figure}

A protohalo is characterized by the mass $M_R$ first collapsing at
the redshift $z$, which is called the instant of halo formation. The
mass $M_R$ corresponds to the linear comoving size $R$:
\be\label{mR}
M_R=\frac{4\pi}{3}\rho_{\rm m} R^{\,3}\,.
\ee
A sphere with the radius $R = 10$ Mpc comprises the typical galaxy
cluster mass $M_{10}\simeq 1.6\cdot 10^{14}\, M_\odot$ (see Section
3).

To single out a protohalo from the common density field, it is
necessary that the protohalo have an initial threshold mass excess
in the sphere of radius $R$ which provides the total negative
binding energy of this region and ensures its collapse by the given
time instant (see Appendix A):
\be\label{por}
\nu_R\equiv\frac{\delta_R}{\sigma_R}\ge\nu\equiv\frac{\delta_c
\left(z\right)}{\sigma_R}\,.
\ee
The function $\delta_c(z)$, which can be calculated analytically, is
controlled numerically. In the CSM, $\delta_c=
\bar{g}(z)\cdot\delta_c(z)={\rm const} \simeq 1.67$, where
$\bar{g}(z)$ is the growth factor of linear density perturbations
($\bar{g}(0)=1$). The initial ensemble of protohalos of interest
here, with initial masses
\be
M_h\ge M_R
\ee
includes only those space points $\mathbf x$ for which the mean
matter density inside a sphere of the radius $R$ around the given
point increases the threshold value in (22), where $\delta
(\mathbf{x}) \equiv\rho(\mathbf{x})/\rho_{\rm m}-1$ is the linear
density contrast relative to the mean density in the Universe
$\rho_{\rm m}$; the smoothed density contrast $\delta_R$ with the
filter of radius $R$ is
\be\label{dR}
\delta_R=\delta_R\!\left(\mathbf{x}\right)=\int \delta\!\left(
\mathbf{x}^\prime\right) W_R\!\left(|\mathbf{x}
-\mathbf{x}^\prime|\right) d\mathbf{x}^\prime \, ,
\ee
where $W_R(r)$ is the `window' function
$$
W_R\!\left(r\right)= \frac{3}{4\pi R^3}\left\{
\begin{array}{l}
1\;,\qquad r\le R\\
0\;,\qquad r>R\;\;
\end{array}
\right.\;\;\;
$$
Clearly, function (24) contains no information on perturbations on
scales smaller than $R$. The variance of the smoothed density linear
contrast $\sigma_R$, linearly approximated to the present moment
$z=0$, is related to the power spectrum of inhomogeneities $P(k)$ by
the integral relation (see Appendix A)
\be\label{sigmR}
\sigma_{\!R}^{\,2}\equiv\langle\delta_R^{\,2}\rangle= \int_0^\infty
\!P\left(k\right) W^2\!\left(kR\right) k^2 dk\,,
\ee
where $W(y)=3 y^{-3}(\sin y -y\cos y)$ is the Fourier transform of
the window function ($y = kR$). It is known from observations that
$\sigma_{11}\simeq 0.8$ for the radius $R = 11$ Mpc, comprising the
galaxy cluster mass $M_{11}\simeq 2\cdot 10^{14}\,M_\odot$. In
astronomy, it is equivalent to the historically established
normalization to `sigma eight', where the index `eight' corresponds
to the physical radius $R = 8/0.7$ Mpc $\simeq 11$ Mpc.

We use the obtained ensembles to describe the mean `thermodynamic'
characteristics of DM in protohalos. The total entropy of a
virialized halo integrates the joint effect of all nonequilibrium
processes that occurred during the entire halo formation time,
including both the background particle (coarse-grained) entropy
related to the initial cosmological perturbations and the acquired
entropy generated during the violent relaxation and hierarchical
clustering of matter. The cosmological part can be calculated
analytically, and to estimate the nonlinear component, we use
numerical results.

\section{From density profile to entropy}

First of all, it should be understood how the entropy is distributed
inside an equilibrium halo. To simplify calculations, we restrict
ourselves by considering the power-law approximation of the central
halo density profile (5):
\be\label{dprof}
\rho(r)\propto r^{-\alpha}\,,
\ee
$$
M(r)=4\pi\int_0^r\rho(x)\, x^2 dx\propto r^{3-\alpha}\,,
$$
with the slope exponent $\alpha\in(0;\, 2.5)$. In numerical
calculations of the halo formation, profiles with $\alpha \gtrsim 1$
emerge, while observations typically show that $\alpha < 1$
(Fig.~10). In the first case, we are dealing with a cusp
($1\le\alpha< 2.5$) and in the second case, with a core ($0
\le\alpha< 1$).

\begin{figure}[t]
\epsfxsize=100 mm \centerline{\epsfbox{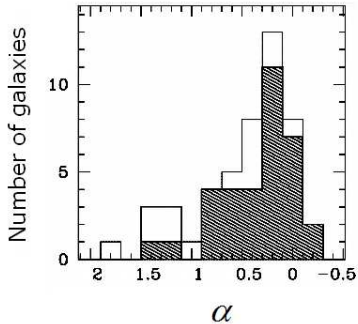}}
\caption{Distribution of the power-law exponent $\alpha$ in the
central part of galaxies with a low surface brightness
(see~\cite{61_deBlok2001}).} \label{ris343}
\end{figure}

The physical difference between the cusp and the core follows from
the behavior of the effective pressure, which can be determined from
Eqn (7):
\be p\!\left(r\right)=c_1+c_2\,r^{2(1-\alpha)}\,,
\ee
where $c_1$ and $c_2$ are integration constants. The critical slope
is $\alpha = 1$: in a core, the pressure is finite, while in a cusp
it diverges at the center.

From Eqn (8), we obtain the entropy mass function
\be\label{beta12}
{\rm E}(M)\,\propto\, c_1\, M^{\beta_1}+c_2\, M^{\beta_2}\,
\propto\, M^\beta\,,
\ee
$$
\beta_1=\frac{5\alpha}{3(3-\alpha)}\,,\qquad \beta_2=\frac{6-
\alpha}{3(3-\alpha)}\,,
$$
where the parameter $\beta=d\ln{\rm E}/d\ln M$ lies in the range
between $\beta_1$ and $\beta_2$ and its local value depends on the
current mass of the halo. At $\alpha =\alpha_c \equiv 1$, the
interval of possible values of $\beta$ contracts to the point
$$
\beta_1= \beta_2= \beta_c\equiv \frac{5}{6}\,.
$$
We have passed from the radius $r$ to the comoving mass $M$ in the
sphere of radius $r$ because this mass is conserved during both
linear perturbations and the halo relaxation.

\begin{figure}[t]
\epsfxsize=100 mm \centerline{\epsfbox{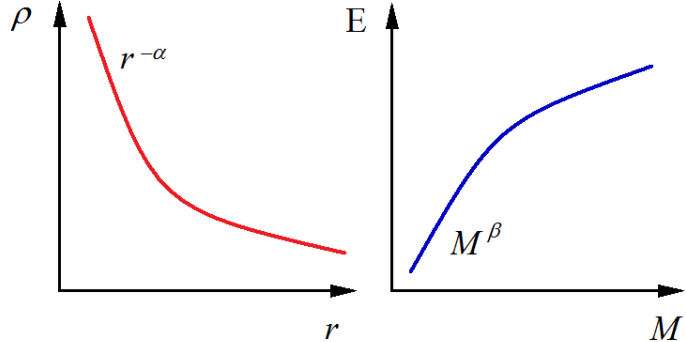}}
\caption{Profiles of (a) the density and (b) the entropy in a DM
halo~\cite{2_Lukash2010}.} \label{ris344}
\end{figure}

As stressed above, the entropy mass function accumulates the action
of irreversible processes during DM evolution and determines the
equilibrium halo density profile. In the considered range of
$\alpha$, we have $\beta_{1,2} > 0$, and hence $E(M) \rightarrow 0$
as $M \rightarrow 0$ (Fig.~11). But in the cusp center, the entropy
of particles is very small,
$$
\beta > 5/6\,:\qquad\;\;\;\; M^{-5/6} \,{\rm E}(M)\rightarrow
0\;\;\;\;\mbox{as}\;\;\;\; M\rightarrow 0\,,
$$
while in the core center it is much higher, which makes the cusp
formation impossible,
$$
\beta < 5/6\,:\qquad\;\;\;\; M^{-5/6} \,{\rm E}(M)\rightarrow \infty
\;\;\;\;\mbox{as}\;\;\;\; M\rightarrow 0\,.
$$

For the critical value $\beta = 5/6$, it follows from (8) and (28)
that
\be\label{TF}
M\propto \sigma^4\,.
\ee
This formula reproduces the well-known empirical
Faber--Jackson~\cite{62_Faber1976} and
Tully--Fisher~\cite{63_Tully1977} relations, which relate the
observed velocity dispersion of stars or gas to the mass
(luminosity) in elliptical galaxies or spherical subsystems of
spiral galaxies.

\begin{figure}[t]
\epsfxsize=100 mm \centerline{\epsfbox{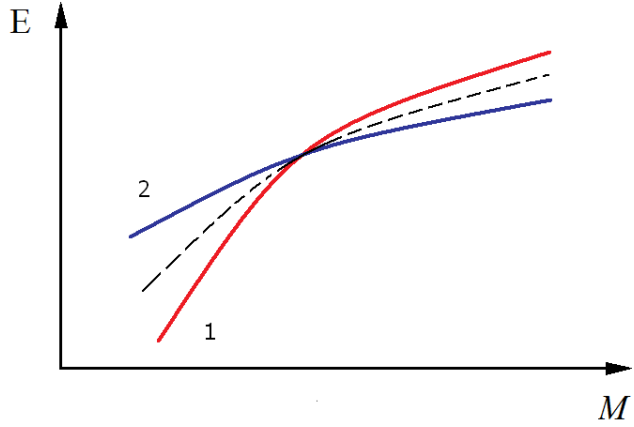}} \caption{The
entropy mass function of an equilibrium halo. Curve $1$ shows the
cusp, curve $2$ shows the core, and the dashed curve corresponds to
$\beta_c = 5/6$.} \label{ris345}
\end{figure}

Figure~12 illustrates the difference between the cusp and the core
of an equilibrium halo from the standpoint of the entropy
distribution. It follows that the formation of the central cusp or
core depends on the total entropy of matter acquired before the
formation of the halo, during and after its formation.

We can now quantitatively answer the question as to how the
initially cold DM particles of the CSM acquire a nonzero
temperature.

The halo formation process can be divided into several consecutive
stages.
\begin{itemize}
\item Initial DM motions are potential and are correlated, relative
velocities of close particles are small, and the DM velocity field
is smooth. Random deviations from the mean velocity determined by
the initial perturbations spectrum increase with increasing the
scale to saturate at relatively large scales ($\sim 40$ Mpc) (see
Fig.~14 in Section 10).
\item During contraction of a protohalo, particles approach each other
and mix at small scales, and the initial phase of the small-scale
part of the random velocity is lost: DM particles heat up (the
background entropy mass function $E_b$
arises).\footnote{Coarse-grained entropy can be defined as the
fraction of the phase space volume per particle:
$$
E_b \propto {\left(\frac{\Delta N}{\Delta {\mathbf x}\Delta {\mathbf
v}}\right)}^{-2/3} \simeq \frac{\sigma^2_{\Delta {\mathbf
v}}}{n^{2/3}}\;,
$$
where $\Delta N$ is the number of particles in the volume $\Delta
{\mathbf x} = \Delta x_1 \Delta x_2 \Delta x_3$, including the
oppositely directed particle flows, $\Delta v_i$ is the relative
particle flow velocity on the scale $\Delta x_i$ ($\Delta {\mathbf
v} = \Delta v_1 \Delta v_2 \Delta v_3$), and $\sigma_{\Delta
{\mathbf v}}$ is the dispersion of relative particle flow velocities
inside the volume $\Delta {\mathbf x}$. The coarse-grained entropy
does not decrease during relaxation of particle flows and is
transformed into the real entropy mass function of the virialized
system.}

\item The chaotization of the regular contraction rate and
transformation of the kinetic energy of the contraction into thermal
energy occur somewhat later during violent relaxation and merger of
clumps (the entropy $E_g$ is generated).
\item The resulting density profile of equilibrium halos is obtained
by adding the initial entropy $E_b$ and the acquired entropy $E_g$.
\end{itemize}

In Sections 10 and 11, we obtain the initial mean entropy profile
$\langle{\rm E}_b \rangle$ using the corresponding dispersion of
linear velocity perturbations [see Section 11, formula (49)]. This
representation assumes an almost adiabatic contraction of matter and
gives the lower limit of the mean function $E_b(M)$ at small $M$.

\section{Initial cosmological perturbations}

The field of scalar cosmological perturbations is characterized by
three gauge-invariant variables: the space vector of the shift of a
material point from the unperturbed position
\be\label{sm}
\mathbf{S}={\mathbf S}\!\left({\mathbf x}\right),
\ee
the total velocity of matter
\be
\mathbf{V}=\mathbf{V}\!\left(z,{\mathbf x}\right),
\ee
and the comoving density perturbation
\be
\delta_{\rm m}\equiv\frac{\rho\!\left(z,\mathbf{x}\right)}{
\rho\!\left(z\right)}-1 = \bar{g}\!\left(z\right)\cdot\delta\!
\left(\mathbf x\right).
\ee
The dependence of the mean density on the redshift has the form
$\,\rho\!\left(z\right)=\rho_{\rm m} \left(1+z\right)^3$, where the
DM density is given by (see Section 3)
$$
\rho_{\rm m}=\frac{3\Omega_{\rm m} H_0^2}{8\pi G}\simeq 3 \cdot
10^{-30}\; \frac{\mbox{g}}{\mbox{sm}^3}\,.
$$
The mean particle shift in the Universe is zero, and its dispersion
is determined by the integral of the power spectrum of density
perturbations:
\be\label{sigm}
\sigma_{\mathbf S}\equiv\sqrt{\langle\mathbf{S}^2\rangle}=\left(
\int_0^\infty\!P(k) dk\right)^{1/2}\!\simeq 14 \;\mbox{Mpc} \,.
\ee
A sphere with the radius $R=\sigma_{\mathbf S}$ contains the mass of
a rich galaxy cluster
$$
M_{\mathbf S}=\frac{4\pi}{3}\rho_{\rm m}\,\sigma_{\mathbf S}^3
\simeq 4\cdot 10^{14}\,M_\odot\,.
$$

Relations between variables are determined by the linear
perturbation theory (see~\cite{1_Lukash2011} for more details):
\be
{\symbolbold{\mathfrak r}}\!\left(z,\mathbf{x}\right) =
\left(1+z\right)^{-1}\left(\mathbf{x} + \bar g \mathbf S\right),
\ee
\be
\mathbf{V}\equiv\dot{\symbolbold{\mathfrak r}} = H \left(
{\symbolbold{\mathfrak r}} - \bar g^{\,\prime}_z \mathbf{S}\right),
\qquad \delta\!\left(\mathbf x\right) = -\,{\rm div} \,\mathbf S\,,
\ee
\be\label{v0}
\mathbf{v}\equiv\mathbf{V} - H{\symbolbold{\mathfrak r}}= -
H\bar{g}^{\,\prime}_z \mathbf{S}\,,\qquad \sigma_{\mathbf{v}}
\equiv\sqrt{\langle\mathbf{v}^2\rangle}\simeq 200 \,
\sqrt{\frac{10}{1+z}}\; \frac{\mbox{km}}{\mbox{s}}\,,
\ee
where ${\symbolbold{\mathfrak r}}(z,\mathbf x)$ and $\mathbf x$ are
the Eulerian and Lagrangian coordinates of the medium, $\mathbf{v}
\equiv\mathbf{v}_{\rm{pec}}$ and $\sigma_{\mathbf v}$ are the
peculiar velocity and its variance, and ${\bar g}^{'}_z$ is the
derivative with respect to the redshift $z$. The growth factor $\bar
g=\bar g(z)$ is normalized to unity at zero redshift ($\bar
g(0)=1$), and at $z > 1$,
$$
\bar g\simeq \frac{1.3}{1+z}\;\;.
$$
The Hubble function and the relation between the redshift and the
peculiar velocity of matter at $z > 1$ have the form
\be\label{smv}
H\simeq 0.5\cdot\left(1+z\right)^{3/2}H_0\,,
\ee
$$
\mathbf{v}\simeq 0.5\cdot H_0\sqrt{1+z}\,\bar{g} \mathbf{S} \simeq
\frac{45}{\sqrt{1+z}}\left(\frac{\mathbf S}{\mbox{Mpc}}\right) \;
\frac{\mbox{km}}{\mbox{s}}\,,
$$

To characterize small-scale motions of matter in a protohalo, we
introduce the vector of the relative shift of medium points
separated by a distance $\mathbf r$ [see (30)]:
\be\label{otnv}
\mathbf{s}=\mathbf{s}\!\left(\mathbf{r},\mathbf{x}\right)=
\frac{\mathbf{S}\!\left(\mathbf{x}+\mathbf{r}\right)-
\mathbf{S}\!\left(\mathbf{x}\right)}{{\sqrt 2}\,\sigma_{\mathbf
S}}\,.
\ee
The vector ${\mathbf s}$ describes the field of the normalized
peculiar velocity of matter with the current radius vector ${\mathbf
r}$ referenced to the point ${\mathbf x}$. We are interested in the
mean relative velocity and its variance during averaging over
different ensembles (or points ${\mathbf x}$) in different regions
of the Universe.

The modulus of the radius vector $r=|\mathbf{r}|$ is fixed in the
interval of linear scales corresponding to inner regions of a
gravitationally bound halo of size $R$:
\be\label{rR}
r < R\,.
\ee
At $R < \sigma_{\mathbf S}$, these scales relate to the
short-wavelength part of the spectrum $P(k)$ (to the right of its
maximum in Fig.~13),
$$
k \sigma_{\mathbf S} \gg 1\,:\qquad\quad
P\!\left(k\right)\propto k^{-3}\ln^2 \!\left(k\sigma_{\mathbf
S}\right)\,.
$$
Hence, we find the estimate for $\sigma_r$ [see (25)]
\be\label{sigmr}
\sigma_r^2=\int_0^\infty \!P\!\left(k\right) W^2\!\left(kr\right)
k^2 dk\propto \ln^3\!\left(2\sigma_{\mathbf S}/r\right).
\ee

\begin{figure}[t]
\epsfxsize=100 mm \centerline{\epsfbox{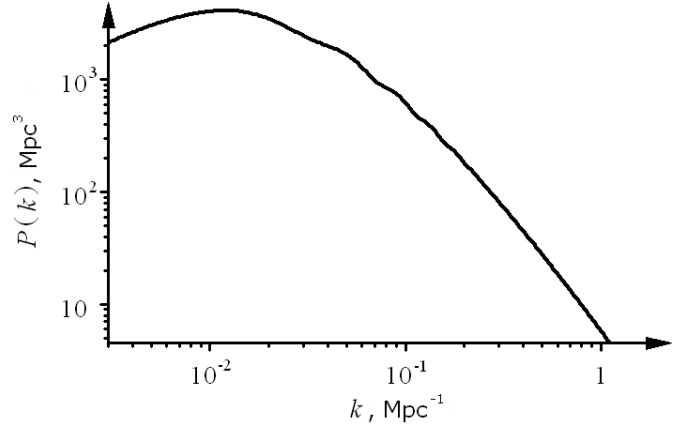}}
\caption{The power spectrum $P(k)$ in the CSM, $\sigma_{10} = 1$.}
\label{amalin_pk}
\end{figure}

Taking the power spectrum normalization $\sigma_{10}\simeq 1$ into
account, we can approximately calculate
$$
1\,\lesssim\,\sigma_R\sim \ln^{3/2}\!\left(2\sigma_{\mathbf S}
/r\right) \qquad \mbox{at} \qquad R\,\lesssim\, 10\; \mbox{Mpc} \,.
$$
At $R > 10$ Mpc, a more accurate estimate for $\sigma_R < 1$ should
be used, taking the change in the spectral slope $P(k)$ at scales of
rich galaxy clusters into account.

Averaging over the entire Universe, we obtain the mean relative
velocity and its variance:
\be
\langle\mathbf{s}\rangle=0\,,
\ee
\be\label{fs}
\sigma_{\mathbf{s}}^2\equiv\langle\mathbf{s}^2\rangle=\sigma_{\mathbf
S}^{ -2}\int_0^\infty \!P\!\left(k\right)\left( 1-\frac{\sin
kr}{kr}\right)dk\,.
\ee
The random part of the velocity
$\sigma_{\mathbf{s}}=\sigma_{\mathbf{s}}(r)$ monotonically increases
with increasing the scale (Fig.~14). At $r < \sigma_{\mathbf{S}}$,
its growth is of a quasi-power form with a smoothly decreasing slope
exponent, starting from unity:
$$
\sigma_{\mathbf{s}}\simeq\frac{r\sigma_r}{\sqrt{6}\,\sigma_{\mathbf
S}}\,.
$$
The growth in the function $\sigma_{ \mathbf{s}}$ decreases at
$r\,\gtrsim\, \sigma_{\mathbf S}$ and saturates at $r\gg
\sigma_{\mathbf S}$: $\sigma_{\mathbf{s}} \rightarrow 1$. At
$r=\sigma_{\mathbf S},\, \sigma_{\mathbf{s}}\simeq 1/2$.

\begin{figure}[t]
\epsfxsize=100 mm \centerline{\epsfbox{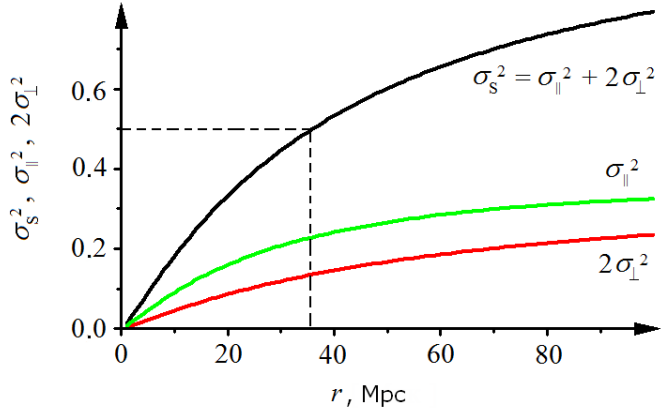}}
\caption{The variance of the relative velocity
$\sigma_{\mathbf{s}}^2(r)$ and its longitudinal $\sigma_{||}^2$ and
transverse $\sigma_{\bot}^2$ components (see~\cite{2_Lukash2010}).}
\label{amalin123}
\end{figure}

The correlation radius of function (42), at which
$\sigma_{\mathbf{s}}^2(r_{\mathbf s})= 1/2$, is $r_{\mathbf s}\sim
36$ Mpc. We note that this is the mean correlation radius of the
relative velocity of matter. The variance of vector (38) is
anisotropic (different for the longitudinal and transverse
components) and depends on the location of particles relative to the
halo center ${\mathbf r}=r{\mathbf e}\,$:
\be\label{fs1} c_{ij}\equiv\langle s_i s_j\rangle=\sigma_{||}^2\,
e_i e_j + \sigma_{\bot}^2\, p_{ij}\,,
\ee
\be\label{fs2} \sigma_{\mathbf s}^2= c_i^i=\sigma_{||}^2 +
2\sigma_{\bot}^2\,,
\ee
where the tensor $ p_{ij}=\delta_{ij} - e_i e_j$ is the projector on
the plane perpendicular to the vector $\mathbf e$,
$$
\sigma_{||}^2= \sigma_{\mathbf S}^{-2} \int_0^\infty\!
P\!\left(k\right) \left[ \frac{1+2W\!\left(kr\right)}{3}-\frac{\sin
kr}{kr}\right] dk\,,
$$
$$
\sigma_{\bot}^2=\frac 13\, \sigma_{\mathbf S}^{-2} \int_0^\infty\!
P\!\left(k\right) \left[1-W\!\left(kr\right)\right] dk\,.
$$
The correlation radii of the longitudinal and transverse variance
are 20 Mpc ($\sigma^2_{||}\simeq 1/6$) and 49 Mpc ($\sigma^2_{\bot}
\simeq 1/6$).

Averaging over the ensemble of protohalos (see Appendix B) with a
density contrast $\nu_R$ [see (22)], we obtain the conditional
expectations of small-scale velocities of matter in a protohalo:
\be\label{sss}
\langle\mathbf{s}|\,\nu_R\rangle= \mathbf{c}\, \nu_R\,,
\ee
\be\label{vars}
\sigma_{\mathbf{s}|\nu}^2\equiv\langle\mathbf{s}^2|\nu_R\rangle -
\langle\mathbf{s}\,|\nu_R\rangle^2 = \sigma_{\mathbf{s}}^2-
\mathbf{c}^2\,,
\ee
where the cross-correlation coefficient has the form [see (39)]
$$
{\mathbf c}\equiv\langle\mathbf{s}\,\nu_R\rangle=-\frac{\mathbf
r}{3\sqrt{2 }\,\sigma_{\mathbf S}\,\sigma_R} \int_0^\infty\!
P\!\left(k\right) W\! \left(kR\right) W\!\left(kr\right)\, k^2 dk
\simeq-\frac{\mathbf{r}\,\sigma_R}{3\sqrt{2}\,\sigma_{\mathbf S}}
\,.
$$
Function (45) determines the universal velocity profile of a
collapsing protohalo with the mass $M_R$. Conditional variance (46)
describes dispersions of random deviations of the velocity from the
mean value inside the protohalo:
\be\label{sigmn}
\sigma_{\mathbf{s}|\nu}^2(r)= \sigma_{\mathbf{s}}^2 - \frac{1}{18}
\left(\frac{r \sigma_R}{\sigma_{\mathbf S}}\right)^2 .
\ee

\section{A solution to the problem of galactic cusps}

The suppression factor of peculiar motions of DM in a protohalo
relative to the motions of matter in the Universe as a whole,
$$
f\!\left(r|R\right)\equiv\frac{\sigma_{\mathbf{s}|\nu}^2}{
\sigma_{\mathbf{s}}^2\,} \simeq 1-\frac{1}{18} \left(\frac{r
\sigma_R}{\sigma_{\mathbf s}\sigma_{\mathbf S}}\right)^2 ,
$$
is shown in Fig.~15 for the galactic mass $1.3\cdot
10^{12}\;M_{\odot}$ ($R = 2$ Mpc). For galaxies and groups of
galaxies, the cross-correlation inside the protohalo is
insignificant:
\be\label{sigmg}
\sigma_{\mathbf{s}|\nu}^2(r) \simeq \frac{r^2}{6\sigma_{\mathbf
S}^2} \left(\sigma_r^2-\frac 13 \sigma_R^2\right)\simeq
\sigma_{\mathbf{s}}^2\,,
\ee
and the variance of small-scale velocities of DM has the universal
form, which is independent of the protohalo mass [see (36), (37)],
$$
\sigma_{\rm v}\equiv\sigma_{\mathbf{v}}\,\sigma_{\mathbf{s}|\nu}
\simeq \sigma_{\mathbf{v}}\,\sigma_{\mathbf{s}}\,,
$$
however, the value of the variance depends on the halo formation
instant $z$.

\begin{figure}[t]
\epsfxsize=100mm \centerline{\epsfbox{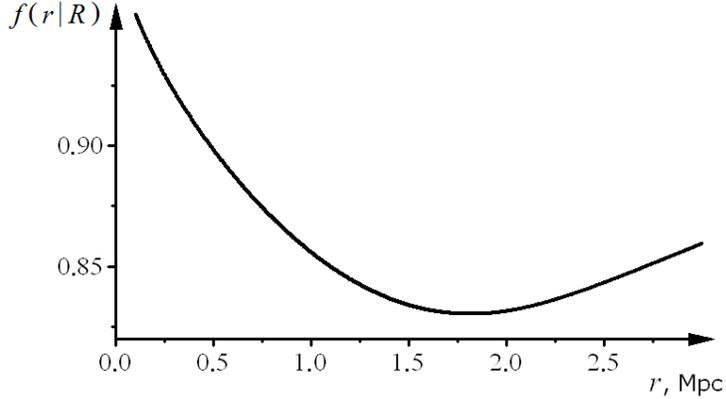}}
\caption{The suppression factor $f\!\left(r|R\right)$ of peculiar
velocities inside a halo with the mass $M=1.3\times 10^{12}\;
M_{\odot}$ (linear size $R = 2$ Mpc) (see~\cite{2_Lukash2010}).}
\label{f7amalin}
\end{figure}

By assuming that the coarse-grained entropy $E(M)$ does not change
in the protohalo contraction (see footnote 3 in Section 9), we
obtain the mean profile of the background entropy mass function (8)
in parametric representation~\cite{2_Lukash2010,41_Mikheeva2007}:
\be\label{sef}
\langle{\rm E}_b\rangle= \frac{\sigma_{{\rm
v}}^2}{3}\left(\frac{m_{\rm p}}{\rho(z)}\right)^{2/3}\! \propto
\sigma_{\mathbf{s}}^2\!\left(r\right)\,, \qquad M=M_r\propto r^3\,.
\ee
Hence, we calculate the power-law slope exponent $\beta_b$ depending
on the index $n$ of the central halo mass $M\equiv 10^n \,M_\odot <
M_R < M_{\mathbf S}\simeq 4 \cdot 10^{14} \,M_\odot$\,:
\be
\beta_b=\frac{d\ln \langle{\rm E}_b\rangle}{d\ln M}=\frac 23-
\frac{1}{\ln(2\sigma_{\mathbf S}/r)}\simeq \frac 23
-\frac{1.3}{16-n}\,,
\ee
%

\begin{table}[t]
\caption{The power-law exponent of the background entropy function
$\beta_b$ for the central halo masses $M\equiv 10^n M_\odot < M_R <
4\cdot 10^{14} M_\odot$.}
\begin{tabular}{cccccc}
\hline
 $\quad n \qquad $ &  $\qquad 12\qquad $ &  $\qquad 10\qquad $ &
 $\qquad 8 \qquad $ &  $\qquad 6 \qquad $ &  $\qquad 4 \quad $ \\
\hline
$\beta_b$ & 0.34 & 0.45 & 0.50 & 0.53 & 0.56\\
\hline \label{tab31}
\end{tabular}
\end{table}

The Table lists numerical values of $\beta_b$ for different $n$. We
see that in the halos of galaxies and groups of galaxies, the
background entropy hampers the central cusp formation, because
\be
\beta_b < 0.6 < \frac 56\,.
\ee
As $R$ increases in the region $R > \sigma_{\mathbf S}$ and the
inner radius $r < R$ increases proportionally (i.e., for large halos
with $M_R \ge 4\cdot 10^{14} M_\odot$),\footnote{We note that the
radius of influence of dark matter $r$ cannot be considered small,
taking the cooling baryon flows and massive cD-galaxies in the
cluster centers into account. For a similar reason, it makes no
sense to consider central DM masses lower than that of black holes
in galaxy centers. We should bear in mind in this connection that
the characteristic scales of DM distributions are much larger than
the sizes of central baryon condensations. } the cross-correlation
$\langle\mathbf{s}\,\nu_R\rangle$ cannot be neglected anymore [see
(47)], and small-scale motions of DM in the protohalo are
suppressed. Therefore, the effect of the background entropy of DM
inside rich galaxy clusters decreases.

Thus, DM cores are formed in the central region of halos of galaxies
and galaxy groups. Following astronomical tradition, these DM cores
can be appropriately called a spherical subsystem or `bulge' of DM
(in analogy with the baryonic bulge). By assuming that $\alpha \sim
1$ in the region with maximum luminosity [see (29)], the
Faber--Jackson relation must be valid for bulges, which is actually
observed in elliptical galaxies. In disk and spiral galaxies, the
bulge occupies a relatively small part of the halo, although its
luminosity in the $H_\alpha$ line can be significant, which
apparently explains the Tully--Fisher relation.

In large spiral galaxies, the DM bulge is deformed due to baryonic
inflow\,\footnote{It should be noted that during a slow baryonic
inflow, the DM cusp (if it exists) would not be destroyed, but only
enhanced. In this sense, it is asymptotically stable. The question
is: what is the DM density in the halo center? This density, in
particular, determines the possible flux of DM annihilation products
(photons, positrons, antiprotons, etc.). Additional studies are
needed to answer this question. Dwarf galaxies with low surface
brightness appear to be ideal objects for studies of the central DM
distribution (see Fig.~3), in which star formation is suppressed and
the gas sustained in rotational equilibrium does not fall onto the
halo center. Studies of spiral galaxies have not provided evidence
of the existence of central cusps, either. Weak gravitational
lensing plays an increasingly important role in reconstructing the
DM surface density.} however, its size is still quite large. For
example, the bulge of our Galaxy occupies several kiloparsecs. As
$r$ increases, it transforms into a DM distribution with
$\rho\propto r^{-2}$ and an almost flat rotation curve, which in
turn extends up to 10 kpc and, possibly, even further beyond the
disk edge; further DM distribution can be probed by the motion of
dwarf satellites.

On the other hand, numerical modeling shows that in gravitationally
bound galaxy groups, the central bulge could be formed only in
compact groups. There is a large class of smoothed galaxy groups
that are only partially relaxed; such is the Local Group of
galaxies. Its mass $\sim 2\cdot 10^{12}\, M_\odot$ is mainly due to
the masses of two large spiral galaxies (the Milky Way and Andromeda
Nebula) and also includes a small addition from masses of several
dozen dwarf galaxies.

When comparing theoretical predictions with observations, a wide
dispersion in the background entropy distribution (and the
corresponding inner density profiles) of relaxed halos, which is
actually observed in galaxies (see Fig.~10) and follows from the CSM
theory, should be taken into account. Indeed, for the initial
Gaussian velocity perturbations, the probability distribution of the
background entropy has the form
\be\label{pdf}
d\, p\!\left(f_b\right)=\frac{\exp\left(-f_b/2\right)}{\sqrt{2\pi
f_b}}\,d f_b\,,
\ee
where
$$
f_b\equiv\frac{{\rm E}_b}{\langle{\rm E}_b\rangle} \simeq
\frac{\mathbf{s}^2}{\sigma_{\mathbf{s}}^2}\,.
$$
We see that the halo background entropy has a wide distribution and
its variances around mean value (49) are large:
$$
\langle f_b\rangle= 1\,,\qquad \langle f_b^2\rangle= 3\langle
f_b\rangle= 3 \,.
$$
These large values
$$
{\rm var}\, {\rm E}_b\, >\, \langle{\rm E}_b\rangle^2
$$
imply significant deviations from the mean power-law slope exponents
of the central density profiles in different galaxy halos, which is
confirmed by observations.

The problem of galactic rotation curves is tightly connected with
the inner structure of equilibrium halos, and we discuss it in
Section 12.

\section{Galactic rotation curves}

In Section 5, we considered flat rotation curves of stellar and gas
disks observed in many galaxies, which suggest an isothermal DM
distribution inside halos in the rage of radii and masses $_\lesssim
10$. The corresponding power-law exponents of the density and
entropy profiles in such systems [$r > r_0$; see (11) and (24)] are
$$
\alpha\simeq 2\,,\qquad \beta\simeq \frac 43\,.
$$
As noted in Section 6, similar distributions are generated in
analytic models of the spherically symmetric or quasispherical
collapse of gravitating dust matter with a smooth initial density
distribution, in which background small-scale perturbations are
definitely absent.During nonlinear contraction of such a cloud,
caustics (self-intersections) and multi-flow motions of matter arise
in its central region, inducing a violent, collisionless relaxation.
As a result of the gravitational redistribution of energy between
spherical shells of matter, power-law density and entropy profiles
are formed in the central parts of the halo, with the exponents
\be\label{gene}
1.6 < \alpha_g < 2\,,\qquad 1 < \beta_g < 1.3\,.
\ee

It follows from (53) that the violent relaxation leads to the
entropy distribution implying the formation of a cusp. However,
these exponents are related to rather large radii. In the central
halo, generated entropy (53) cannot compete with the background one
and turns out to be negligibly small (see the Table). Therefore, the
background entropy effect constrains the central density and hinders
the formation of cusp-like halo profiles.

The same conclusion is valid for the numerical experiment modeling
of the most complex stages of halo formation that cannot be treated
analytically. Violent relaxation, in which radial trajectories play
the dominant role, is not the only and, possibly, the main
relaxation process of collisionless systems that decoupled from the
Friedman flow during cosmological expansion. An important role here
is played by anisotropic collapse and the effects of merger of
clumps with different masses that were formed before the central
massive part and simultaneously with it. All these processes, which
are actually observed in nonlinearly relaxing gravitationally bound
systems at high redshifts, are called hierarchical clustering. They
include the tidal binary (and also triple, etc.) interaction of
clumps of matter captured in the common gravitational field and
their scattering and merging into more massive objects with the
subsequent repetition of the entire cycle.

The development of numerical modeling techniques for a large number
of gravitationally interacting bodies has led to the notion of the
universal halo density NFW profile, which appears after averaging
the density profiles in several hundred clumps of different masses
and sizes formed during the merger. The initial particle
distribution in the N-body problem is taken to be close to
cosmological conditions, but because the number N is finite, the
dynamical range of model scales is restricted. As noted in Section
6, the NFW profile asymptotically approaches a cusp-like power law
with the exponents
$$
\alpha_g=1\;\;,\qquad \beta_g=\frac 56\;\;.
$$

In numerical modeling of the $N$-body problem, the background
entropy of small-scale perturbations is apparently underestimated
due to the restricted spatial resolution of this method. Hopefully,
as the power of numerical methods increases, the background entropy
effects can be properly taken into account.

Already much evidence has appeared showing that with increasing $N$,
the exponent $\alpha$ decreases below unity; however, these are
single examples and no statistics can be applied. Due to the
small-scale cut-off of the model power spectrum, the initial entropy
effect is suppressed in comparison with the analytic estimates shown
in the Table.

Of course, the background entropy effect on the internal halo
structure and the formation of the central core with finite pressure
and DM density should be confirmed by numerical experiment.
Nevertheless, today we can already verify the CSM predictions by
measuring velocities of gas and stars in galaxies and galaxy
clusters.

Figure~3 demonstrates galactic rotation curves that cannot be
described by formula (19) but are well approximated by a smoother
profile in (18). Such examples are typical for galaxies (see,
e.g.,~\cite{64_Evans2009}). In the central region with $r < r_s$,
the fit constructed using the observational points is similar to the
isothermal sphere approximation [see (16) and (17)]; but at large
radii, the density decrease is inversely proportional to $r^3$ and
circular velocities slowly decrease. The circular velocity reaches a
maximum at the point of the local density profile with the exponent
$\alpha = 2$.

Simulated NFW profiles meet with another problem: they do not
reproduce flat galactic rotation curves in the observed range of
scales. Nevertheless, numerical profiles provide a good
approximation of the mass distribution in galaxy clusters (see
Fig.~4, 5).

\newpage

\section{Effect of the background entropy}

To understand the physical meaning and variety of the observed
rotation curves, the initial background entropy of DM in nonlinear
halos should be taken into account.

To clarify the physics of this effect, we consider the simplest
model of the joint entropy mass function in the form
\be\label{mod}
{\rm E}(M) = \sqrt{{\rm E}_b^2+{\rm E}_g^2}\;,\qquad {\rm f}_b=
\frac{{\rm E}_b}{\rm E}\;,
\ee
where the parameter ${\rm f}_b\in \left(0, 1\right)$ expresses the
relative contribution of the background entropy at a fixed inner
halo radius corresponding to the circular velocity maximum or the
local density profile slope $\alpha = 2$. The functions ${\rm
E}_{b\,,\,g}(M)$ have a power-law form with the corresponding slope
exponents in the ranges
\be\label{inter}
\beta_b\in \left(\frac 13\,, \;\frac 23\right)\,,\qquad \beta_g\in
\left(\frac 56\,, \;\frac 43\right)\,.
\ee

We now can calculate the rotation curves of particles in equilibrium
DM halos analytically and compare them with observations and
numerical model calculations. The results are presented in Fig.~16.
The rotation velocities $\rm{v}_{\rm rot}=\rm{v}\!\left(r\right)$ as
functions of the inner radius $r$ are obtained for models of
hierarchical ($\beta_g=5/6$, Fig.~16a) and violent ($\beta_g=4/3$,
Fig.~16b) relaxations for two limit values of ${\rm f}_b$, ${\rm
f}_b\ll 1$ and ${\rm f}_b\sim 1$, and three values of the background
entropy indices. Velocities and radii are normalized to the
corresponding maximum values of the velocity and the radius $r_{-2}$
at which it is attained: $\rm{v}_{\rm max}
\!\equiv\rm{v}(r_{-2})\geq \rm{v}(r)$. We see from this figure that
for halos with $\beta_b < 0.5$ and any parameters $\beta_g$ and
${\rm f}_b$ in the chosen intervals, the induced rotation curves
fully cover the range between the NFW and Burkert approximations.
For ${\rm f}_b\ll 1$, the curves shift to the NFW region, while as
${\rm f}_b$ increases, they become similar to the Burkert profiles.

\begin{figure}[t]
\epsfxsize=100 mm \centerline{\epsfbox{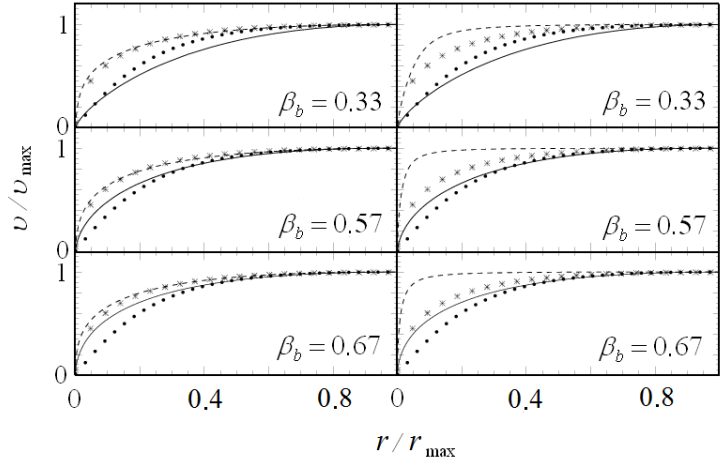}}
\caption{Normalized rotation curves $v\!\left(r\right)$ for models
of (a) hierarchical ($\beta_g =5/6$) and (b) violent ($\beta_g=4/3$)
relaxation at ${\rm f}_b\ll 1$ (the dashed curve) and ${\rm
f}_b\simeq 1$ (the solid curve) (see~\cite{41_Mikheeva2007}). The
NFW and Burkert approximations are respectively shown by X's and the
dotted line.}
\end{figure}

It follows that the observed rotation curves are successfully
reproduced by simple models (54) with the corresponding fitting
parameters ${\rm f}_b$ and $\beta_b$. Distributions of the
background (52) and generated (55) entropies in the halo allow
explaining the required diversity of the observed galactic rotation
curves. As the exponent $\beta_b$ increases, the rotation curves
concentrate closer to the NFW profiles.

The examples considered here prove the following.
\begin{itemize}
\item The background entropy prevents the formation of central
DM cusps in halos in the mass range $10^8-10^{14}\, M_\odot$. For
heavier and lighter halos, the background entropy effect is
attenuated.
\item Taking the effect of the background entropy on the density
distribution inside a halo into account allows reproducing the
rotation curves in a wide range of scales and can solve the problem
of DM cusps in the CSM framework.
\end{itemize}

To conclude, we note that the models considered here are based on
simple assumptions of the DM particle distribution in the halo and
neglect the contribution of baryonic matter. The full problem of the
evolution of galactic density profiles, including DM and baryons and
accounting for the effect of different dissipative processes on the
angular momentum transfer from the central parts of galaxies, is
undoubtedly very complicated and requires further studies, both
analytic and numerical.

\section{Conclusion}

The entropy method of description of virialized DM systems
considered here allows an analytic treatment of these complex
nonlinear structures. It enables connecting the inner DM density
profiles with characteristics of both the initial small-scale field
of density perturbations and the nonlinear large-scale relaxation of
gravitationally contracted matter. We conclude that cosmological
random motions of matter `warm up' DM particles in the collapsing
protohalos. We have shown that taking this effect into account
\begin{itemize}
\item leads to suppression of cusp-like density profiles in the
forming halo and to the formation of DM cores in galaxies;
\item allows explaining the diversity of galactic rotation curves,
both observed and obtained in numerical simulations;
\item helps to solve the problem of the inner structure of equilibrium
DM halos in the CSM framework.
\end{itemize}

The obtained analytic results should be confirmed by numerical
$N$-body experiments, which can be done in the future with improved
spatial resolution of the central halo parts.

The authors thank S V Pilipenko for the fruitful discussions.

This work was supported by the Russian Foundation for Basic Research
grants 11-02-12168-ofi-m-2011 and 11-02-00244, and by the FCP
program Scientific and Scientific--Pedagogical Specialists of
Innovative Russia' for 2009-2013 (state contracts P1336 and
16.740.11.0460).

\section{Appendices}

\subsection*{Appendix A. Correlation functions of the initial perturbations}

Fourier harmonics of the linear density perturbation field are
$\delta$-correlated:
$$
\delta(\mathbf{x})=(2\pi)^{-3/2}\int \delta_{\mathbf
k}\,e^{i{\mathbf k}{ \mathbf x}}d\mathbf{k}\,, \qquad
\langle\delta_{\mathbf k}\delta_{ \mathbf{k}^\prime}^*\rangle=2\pi^2
P(k)\,\delta^{(3)}\!\left(\mathbf{k} - \mathbf{k}^\prime\right),
$$
where the averaging $\langle\ldots\rangle$ is done over all
realizations of the random Gaussian field, $\delta^{(3)}(\mathbf k)$
is the three-dimensional Dirac $\delta$-function, and $P = P(k)$ is
the power spectrum.

Using this decomposition, we can construct any correlators of the
cosmological perturbation field, for example
$$
\langle\delta\!\left( \mathbf{x} + \mathbf{r} \right) \delta\!\left(
\mathbf{x} \right) \rangle = \int_0^\infty\!P(k) \frac{\sin
kr}{kr}\, k^2 dk\,,
$$
$$
\langle \mathbf{S}\!\left(\mathbf{x}
+\mathbf{r}\right)\delta\!\left(\mathbf x\right) \rangle = -
\frac{\mathbf r}{3} \int_0^\infty\! P(k) W(kr)\, k^2 dk\,,
$$
\begin{multline*}
\langle S_j\!\left(\mathbf{x}+\mathbf{r}\right)
S_k\!\left(\mathbf{x}\right) \rangle = e_j e_k \int_0^\infty
\!P\!\left(k\right)\left[\frac{\sin kr}{kr} -\frac 23 W(kr)\right]\!
dk + {} \\
{} + \frac 13\, p_{jk} \int_0^\infty\! P\!\left(k\right)
W\!\left(kr\right) dk
\end{multline*}
where $\mathbf{e}=e_j=r_j/r$ is the unit vector in the direction
$\mathbf r$, $p_{jk}$ is the projection tensor [see (43)], $\delta=
- {\rm div}\, {\mathbf S}$, and ${\mathbf S}$ is the shift vector of
matter elements [see (30)-(37)],
$$\mathbf{S}= S_j= i (2\pi)^{-3/2}\!
\int\frac{k_j}{k^2}\,\delta_{\mathbf k}\,e^{i{\mathbf k}{\mathbf
x}}d\mathbf k
$$
We can represent smoothed fields similarly, for example,
$$
\delta_R=(2\pi)^{-3/2}\!\int \delta_{\mathbf k}\, W\!\left(kR\right)
e^{i{\mathbf k}{\mathbf x}} d\mathbf{k}\,,
$$
$$
\langle \mathbf{S}\!\left(\mathbf{x}
+\mathbf{r}\right)\delta_R\!\left( \mathbf{x}\right)\rangle = -
\frac{\mathbf r}{3} \int_0^\infty\! P\! \left(k\right)
W\!\left(kR\right) W\!\left(kr\right) k^2 dk\,,
$$
etc.

\subsection*{Appendix B. Conditional probability distribution}

To obtain the conditional probability $p({\mathbf s}|\nu_R)$ of
$\mathbf s$ for a given value of $\nu_R$, we consider the Gaussian
distribution of two variables $x_A=(\mathbf{s}, \nu_R)$ with the
index $A$ ranging over the values $i, R$:
$$
p\left(\mathbf{s},\nu_R\right)=\frac{\exp\left(-\kappa^2/2\right)}{
\left(2\pi\right)^2\sqrt C}\,, \qquad \kappa^2= c^{\,AB} x_A x_B\,,
$$
where $c^{\,AB}$ is the inverse matrix to the matrix [see (43),
(45)]
$$
c_{AB}=\langle x_A x_B\rangle=\left(
\begin{array}{cc}
c_{ij} & \mathbf{c} \\
\mathbf{c} & 1\\
\end{array}
\right),
$$
$$
C\equiv\det \left(c_{AB}\right)= \bar{\sigma}_{||}^{\,2}
\,\sigma_{\bot}^4\,,\qquad \bar{\sigma}_{||}^{\,2}=
\sigma_{||}^2-\mathbf{c}^{\,2}\,.
$$
After simple transformations, we obtain
$$
\kappa^2=\bar\kappa^{\,2} +\nu_R^{\,2}\,,\qquad
\bar\kappa^{\,2}=\bar{c}^{\,ij} \bar{s}_i \bar{s}_j\,, \qquad
\bar{\mathbf s}={\mathbf s} - {\mathbf c}\,\nu_R\,,
$$
where $\bar{c}^{\,ij}$ is the matrix inverse to
$$
\bar{c}_{ij}\equiv\langle\bar{s}_i \bar{s}_j|\nu_R\rangle=\bar{
\sigma}_{||}^{\,2} \,e_i e_j + \sigma_{\bot}^{2}\, p_{ij}\,.
$$
Hence, we find the conditional probability distribution of the
vector ${\mathbf s}$ and the variance of velocities inside a halo
with the density contrast $\nu_R$:
$$
p\!\left({\mathbf{s}}|\nu_R\right)=\frac{p\!\left(\mathbf{s},\nu_R
\right)}{p\!\left(\nu_R\right)}=\frac{\exp\left(-
\bar\kappa^{\,2}/2\right)}{\left(2 \pi\right)^{3/2}\bar{\sigma_{||}}
\sigma_\bot^2}\,, $$

$$ \sigma_{\vec{s}|\nu}^2\equiv\langle
\bar{\mathbf{s}}^{\,2}|\nu_R\rangle=\sigma_{\mathbf{
s}}^2-\mathbf{c}^{\,2}\,.
$$

\end{document}